\documentclass[twocolumn,trackchanges,floatfix]{aastex631}

\newcommand{\hess}{{H.E.S.S.}}
\newcommand{\fermi}{\textit{Fermi}}
\newcommand{\swift}{\textit{Swift}}
\newcommand{\nustar}{NuSTAR}
\newcommand{\g}{$\gamma$-ray}
\newcommand{\gs}{$\gamma$-rays}
\newcommand{\src}{PKS~0735+178}
\newcommand{\neu}{IceCube-211208A}

\newcommand{\mpo}{\textcolor{black}}
\newcommand{\mponn}{\textcolor{black}}
\newcommand{\mpon}{\textcolor{black}}
\newcommand{\mc}{\textcolor{black}}
\newcommand{\qf}{\textcolor{black}}
\definecolor{darkspringgreen}{rgb}{0.09, 0.59, 0.16}
\newcommand{\qfn}{\textcolor{black}}
\newcommand{\revA}{\textcolor{black}}
\newcommand{\revHESS}{\textcolor{black}} 

\newcommand{\revB}{\textcolor{black}}
\newcommand{\revC}{\textcolor{black}}
\newcommand{\revHESSc}{\textcolor{black}} 

\newcommand{\revD}{\textcolor{black}}

\accepted{June 30, 2023}

\usepackage{subfigure}
\usepackage{amsmath,multirow, url,hyperref}
\usepackage[normalem]{ulem}
 \usepackage{booktabs}
\submitjournal{ApJ}

\shorttitle{MWL Observations of the Blazar PKS~0735+178/IC211208A}
\shortauthors{The VERITAS and H.E.S.S. Collaborations et al.}
\graphicspath{{./}{figures/}}

\begin{document}

\title{Multiwavelength Observations of the Blazar PKS~0735+178 in Spatial and Temporal Coincidence with an Astrophysical Neutrino Candidate IceCube-211208A}

\correspondingauthor{Qi Feng}
\email{qi.feng@cfa.harvard.edu}
\correspondingauthor{Martin Pohl}
\email{martin.pohl@desy.de}
\correspondingauthor{Matteo Cerruti}
\email{contact.hess@hess-experiment.eu}
\correspondingauthor{Federica Bradascio}
\email{contact.hess@hess-experiment.eu}
\correspondingauthor{Jooyun Woo}
\email{jw3855@columbia.edu}
\correspondingauthor{Atreya Acharyya}
\email{aacharyya1@ua.edu}
\correspondingauthor{Weidong Jin}
\email{wjin4@crimson.ua.edu}
\correspondingauthor{Jean Damascene Mbarubucyeye}
\email{contact.hess@hess-experiment.eu}

\author[0000-0002-2028-9230]{A.~Acharyya}\affiliation{Department of Physics and Astronomy, University of Alabama, Tuscaloosa, AL 35487, USA}

\author[0000-0002-9021-6192]{C.~B.~Adams}\affiliation{Physics Department, Columbia University, New York, NY 10027, USA}
\author{A.~Archer}\affiliation{Department of Physics and Astronomy, DePauw University, Greencastle, IN 46135-0037, USA}
\author[0000-0002-3886-3739]{P.~Bangale}\affiliation{Department of Physics and Astronomy and the Bartol Research Institute, University of Delaware, Newark, DE 19716, USA}
\author[0000-0002-9675-7328]{J.~T.~Bartkoske}\affiliation{Department of Physics and Astronomy, University of Utah, Salt Lake City, UT 84112, USA}
\author{P.~Batista}\affiliation{DESY, Platanenallee 6, 15738 Zeuthen, Germany}
\author[0000-0003-2098-170X]{W.~Benbow}\affiliation{Center for Astrophysics $|$ Harvard \& Smithsonian, Cambridge, MA 02138, USA}
\author[0000-0002-6208-5244]{A.~Brill}\affiliation{N.A.S.A./Goddard Space-Flight Center, Code 661, Greenbelt, MD 20771, USA}
\author[0000-0001-6391-9661]{J.~H.~Buckley}\affiliation{Department of Physics, Washington University, St. Louis, MO 63130, USA}
\author{J.~L.~Christiansen}\affiliation{Physics Department, California Polytechnic State University, San Luis Obispo, CA 94307, USA}
\author{A.~J.~Chromey}\affiliation{Center for Astrophysics $|$ Harvard \& Smithsonian, Cambridge, MA 02138, USA}
\author[0000-0002-1853-863X]{M.~Errando}\affiliation{Department of Physics, Washington University, St. Louis, MO 63130, USA}
\author[0000-0002-5068-7344]{A.~Falcone}\affiliation{Department of Astronomy and Astrophysics, 525 Davey Lab, Pennsylvania State University, University Park, PA 16802, USA}
\author[0000-0001-6674-4238]{Q.~Feng}\affiliation{Center for Astrophysics $|$ Harvard \& Smithsonian, Cambridge, MA 02138, USA}

\author[0000-0002-2944-6060]{G.~M.~Foote}\affiliation{Department of Physics and Astronomy and the Bartol Research Institute, University of Delaware, Newark, DE 19716, USA}
\author[0000-0002-1067-8558]{L.~Fortson}\affiliation{School of Physics and Astronomy, University of Minnesota, Minneapolis, MN 55455, USA}
\author[0000-0003-1614-1273]{A.~Furniss}\affiliation{Department of Physics, California State University - East Bay, Hayward, CA 94542, USA}
\author{G.~Gallagher}\affiliation{Department of Physics and Astronomy, Ball State University, Muncie, IN 47306, USA}
\author[0000-0002-0109-4737]{W.~Hanlon}\affiliation{Center for Astrophysics $|$ Harvard \& Smithsonian, Cambridge, MA 02138, USA}
\author[0000-0002-8513-5603]{D.~Hanna}\affiliation{Physics Department, McGill University, Montreal, QC H3A 2T8, Canada}
\author[0000-0003-3878-1677]{O.~Hervet}\affiliation{Santa Cruz Institute for Particle Physics and Department of Physics, University of California, Santa Cruz, CA 95064, USA}
\author[0000-0001-6951-2299]{C.~E.~Hinrichs}\affiliation{Center for Astrophysics $|$ Harvard \& Smithsonian, Cambridge, MA 02138, USA and Department of Physics and Astronomy, Dartmouth College, 6127 Wilder Laboratory, Hanover, NH 03755 USA}
\author{J.~Hoang}\affiliation{Santa Cruz Institute for Particle Physics and Department of Physics, University of California, Santa Cruz, CA 95064, USA}
\author[0000-0002-6833-0474]{J.~Holder}\affiliation{Department of Physics and Astronomy and the Bartol Research Institute, University of Delaware, Newark, DE 19716, USA}
\author[0000-0002-1432-7771]{T.~B.~Humensky}\affiliation{Department of Physics, University of Maryland, College Park, MD, USA and NASA GSFC, Greenbelt, MD 20771, USA}
\author[0000-0002-1089-1754]{W.~Jin}\affiliation{Department of Physics and Astronomy, University of Alabama, Tuscaloosa, AL 35487, USA}

\author[0000-0002-3638-0637]{P.~Kaaret}\affiliation{Department of Physics and Astronomy, University of Iowa, Van Allen Hall, Iowa City, IA 52242, USA}
\author{M.~Kertzman}\affiliation{Department of Physics and Astronomy, DePauw University, Greencastle, IN 46135-0037, USA}
\author{M.~Kherlakian}\affiliation{DESY, Platanenallee 6, 15738 Zeuthen, Germany}
\author[0000-0003-4785-0101]{D.~Kieda}\affiliation{Department of Physics and Astronomy, University of Utah, Salt Lake City, UT 84112, USA}
\author[0000-0002-4260-9186]{T.~K.~Kleiner}\affiliation{DESY, Platanenallee 6, 15738 Zeuthen, Germany}
\author[0000-0002-4289-7106]{N.~Korzoun}\affiliation{Department of Physics and Astronomy and the Bartol Research Institute, University of Delaware, Newark, DE 19716, USA}
\author[0000-0002-5167-1221]{S.~Kumar}\affiliation{Department of Physics, University of Maryland, College Park, MD, USA }
\author[0000-0003-4641-4201]{M.~J.~Lang}\affiliation{School of Natural Sciences, University of Galway, University Road, Galway, H91 TK33, Ireland}
\author[0000-0003-3802-1619]{M.~Lundy}\affiliation{Physics Department, McGill University, Montreal, QC H3A 2T8, Canada}
\author[0000-0001-9868-4700]{G.~Maier}\affiliation{DESY, Platanenallee 6, 15738 Zeuthen, Germany}
\author{C.~E~McGrath}\affiliation{School of Physics, University College Dublin, Belfield, Dublin 4, Ireland}
\author[0000-0001-7106-8502]{M.~J.~Millard}\affiliation{Department of Physics and Astronomy, University of Iowa, Van Allen Hall, Iowa City, IA 52242, USA}
\author{J.~Millis}\affiliation{Department of Physics and Astronomy, Ball State University, Muncie, IN 47306, USA}
\author[0000-0001-5937-446X]{C.~L.~Mooney}\affiliation{Department of Physics and Astronomy and the Bartol Research Institute, University of Delaware, Newark, DE 19716, USA}
\author[0000-0002-1499-2667]{P.~Moriarty}\affiliation{School of Natural Sciences, University of Galway, University Road, Galway, H91 TK33, Ireland}
\author[0000-0002-3223-0754]{R.~Mukherjee}\affiliation{Department of Physics and Astronomy, Barnard College, Columbia University, NY 10027, USA}
\author[0000-0002-9296-2981]{S.~O'Brien}\affiliation{Physics Department, McGill University, Montreal, QC H3A 2T8, Canada}
\author[0000-0002-4837-5253]{R.~A.~Ong}\affiliation{Department of Physics and Astronomy, University of California, Los Angeles, CA 90095, USA}
\author[0000-0001-7861-1707]{M.~Pohl}\affiliation{Institute of Physics and Astronomy, University of Potsdam, 14476 Potsdam-Golm, Germany and DESY, Platanenallee 6, 15738 Zeuthen, Germany}

\author[0000-0002-0529-1973]{E.~Pueschel}\affiliation{DESY, Platanenallee 6, 15738 Zeuthen, Germany}
\author[0000-0002-4855-2694]{J.~Quinn}\affiliation{School of Physics, University College Dublin, Belfield, Dublin 4, Ireland}
\author[0000-0002-5351-3323]{K.~Ragan}\affiliation{Physics Department, McGill University, Montreal, QC H3A 2T8, Canada}
\author{P.~T.~Reynolds}\affiliation{Department of Physical Sciences, Munster Technological University, Bishopstown, Cork, T12 P928, Ireland}
\author[0000-0002-7523-7366]{D.~Ribeiro}\affiliation{School of Physics and Astronomy, University of Minnesota, Minneapolis, MN 55455, USA}
\author{E.~Roache}\affiliation{Center for Astrophysics $|$ Harvard \& Smithsonian, Cambridge, MA 02138, USA}
\author[0000-0003-1387-8915]{I.~Sadeh}\affiliation{DESY, Platanenallee 6, 15738 Zeuthen, Germany}
\author{A.~C.~Sadun}\affiliation{Campus Box 157, P.O.Box 173364, Denver CO 80217, USA}
\author[0000-0002-3171-5039]{L.~Saha}\affiliation{Center for Astrophysics $|$ Harvard \& Smithsonian, Cambridge, MA 02138, USA}
\author{M.~Santander}\affiliation{Department of Physics and Astronomy, University of Alabama, Tuscaloosa, AL 35487, USA}
\author{G.~H.~Sembroski}\affiliation{Department of Physics and Astronomy, Purdue University, West Lafayette, IN 47907, USA}
\author[0000-0002-9856-989X]{R.~Shang}\affiliation{Department of Physics and Astronomy, Barnard College, Columbia University, NY 10027, USA}
\author{M.~Splettstoesser}\affiliation{Santa Cruz Institute for Particle Physics and Department of Physics, University of California, Santa Cruz, CA 95064, USA}
\author{A.~Kaushik~Talluri}\affiliation{School of Physics and Astronomy, University of Minnesota, Minneapolis, MN 55455, USA}
\author{J.~V.~Tucci}\affiliation{Department of Physics, Indiana University-Purdue University Indianapolis, Indianapolis, IN 46202, USA}
\author{V.~V.~Vassiliev}\affiliation{Department of Physics and Astronomy, University of California, Los Angeles, CA 90095, USA}
\author{A.~Weinstein}\affiliation{Department of Physics and Astronomy, Iowa State University, Ames, IA 50011, USA}
\author[0000-0003-2740-9714]{D.~A.~Williams}\affiliation{Santa Cruz Institute for Particle Physics and Department of Physics, University of California, Santa Cruz, CA 95064, USA}
\author[0000-0002-2730-2733]{S.~L.~Wong}\affiliation{Physics Department, McGill University, Montreal, QC H3A 2T8, Canada}
\author{J.~Woo}\affiliation{Columbia Astrophysics Laboratory, Columbia University, New York, NY 10027, USA}

\collaboration{0}{The VERITAS Collaboration}

\author{F.~Aharonian}
\affiliation{Dublin Institute for Advanced Studies, 31 Fitzwilliam Place, Dublin 2, Ireland}
\affiliation{Max-Planck-Institut f\"ur Kernphysik, P.O. Box 103980, D 69029 Heidelberg, Germany}

\author{J.~Aschersleben}
\affiliation{Kapteyn Astronomical Institute, University of Groningen, Landleven 12, 9747 AD Groningen, The Netherlands}

\author[0000-0002-9326-6400]{M.~Backes}
\affiliation{University of Namibia, Department of Physics, Private Bag 13301, Windhoek 10005, Namibia}
\affiliation{Centre for Space Research, North-West University, Potchefstroom 2520, South Africa}

\author[0000-0002-5085-8828]{V.~Barbosa~Martins}
\affiliation{DESY, D-15738 Zeuthen, Germany}

\author[0000-0002-5797-3386]{R.~Batzofin}
\affiliation{Institut f\"ur Physik und Astronomie, Universit\"at Potsdam,  Karl-Liebknecht-Strasse 24/25, D 14476 Potsdam, Germany}

\author[0000-0002-2115-2930]{Y.~Becherini}
\affiliation{Université de Paris, CNRS, Astroparticule et Cosmologie, F-75013 Paris, France}
\affiliation{Department of Physics and Electrical Engineering, Linnaeus University,  351 95 V\"axj\"o, Sweden}

\author[0000-0002-2918-1824]{D.~Berge}
\affiliation{DESY, D-15738 Zeuthen, Germany}
\affiliation{Institut f\"ur Physik, Humboldt-Universit\"at zu Berlin, Newtonstr. 15, D 12489 Berlin, Germany}

\author[0000-0001-8065-3252]{K.~Bernl\"ohr}
\affiliation{Max-Planck-Institut f\"ur Kernphysik, P.O. Box 103980, D 69029 Heidelberg, Germany}

\author{B.~Bi}
\affiliation{Institut f\"ur Astronomie und Astrophysik, Universit\"at T\"ubingen, Sand 1, D 72076 T\"ubingen, Germany}

\author[0000-0002-8434-5692]{M.~B\"ottcher}
\affiliation{Centre for Space Research, North-West University, Potchefstroom 2520, South Africa}

% ...

\collaboration{0}{The H.E.S.S. Collaboration}

\author{K.~Mori}
\affiliation{Physics Department, Columbia University, New York, NY 10027, USA}
\nocollaboration{1}

\begin{abstract}

We report on multiwavelength target-of-opportunity observations of the blazar PKS~0735+178, located 2.2$^\circ$ away from the best-fit position of the IceCube neutrino event \neu{} detected on December 8, 2021. The source was in a high-flux state in the optical, ultraviolet, X-ray, and GeV \g{} bands around the time of the neutrino event, exhibiting daily variability in the soft X-ray flux. The X-ray data from \swift{}-XRT and \nustar{} characterize the transition between the low-energy and high-energy components of the broadband spectral energy distribution (SED), and the \g{} data from \fermi{}-LAT, VERITAS, and \hess{} require a spectral cut-off near 100 GeV. Both X-ray and \g{} measurements provide strong constraints on the leptonic and hadronic models. We analytically explore a synchrotron self-Compton model, an external Compton model, and a lepto-hadronic model. Models that are entirely based on internal photon fields face serious difficulties in matching the observed SED. The existence of an external photon field in the source would instead explain the observed \g{} spectral cut-off in both leptonic and lepto-hadronic models and allow a proton jet power that marginally agrees with the Eddington limit in the lepto-hadronic model. We show a numerical lepto-hadronic model with external target photons that reproduces the observed SED and is reasonably consistent with the neutrino event despite requiring a  high jet power. 

\end{abstract}



\section{Introduction} \label{sec:intro}

The IceCube Neutrino Observatory has detected a diffuse flux of astrophysical neutrinos \citep{IceCube13}, \revB{whose isotropic distribution of arrival directions suggests an extragalactic origin}, but has not firmly identified any neutrino point sources to date despite strong evidence of TeV neutrino emission from a nearby active galaxy \revB{NGC~1068} 
\citep{Aartsen20,IceCube2022_NGC1068}. A number of extragalactic sources are proposed as candidates for high-energy neutrino emitters, including clusters of galaxies, active galactic nuclei (including blazars), starburst galaxies, $\gamma$-ray bursts, supernovae, \revHESSc{and tidal disruption events \citep[e.g.,][]{Kurahashi2022}.} Any detection of a TeV -- PeV neutrino-emitting source will directly constrain the century-old puzzle of the origin of cosmic rays \citep[e.g.,][]{Meszaros17}, as these neutrinos must be produced by \revA{hadronic} cosmic-ray interactions. 

\revA{Hadronic} cosmic-ray interactions produce not only neutrinos but also $\gamma$-rays (from neutral pion decays) and X-rays (from synchrotron radiation and cascading of the secondary electrons/positrons). Observations at \g{} and X-ray bands are therefore critical for studies of neutrinos and cosmic rays. 
An important type of source that exhibits strong and highly variable \g{} and X-ray emission is blazars, a subclass of active galactic nuclei with relativistic jets pointing \revHESSc{towards} Earth. 
\revB{High-energy} emission from blazars can be produced by either leptonic (via inverse-Compton scattering of relativistic electrons) or hadronic interactions \citep{Bottcher13}, rendering electromagnetic observations, by themselves, generally insufficient to probe the origin of cosmic rays. 

\revA{A more efficient approach to identify astrophysical neutrino and ultra-high-energy cosmic-ray sources is through multi-messenger observations of correlated neutrino and $\gamma$-ray events.}
\revA{H}igh-power $\gamma$-ray blazars are disfavored as the dominant origin of the observed IceCube neutrinos \citep[e.g.,][]{Aartsen17}, \revA{but} individual flaring blazars still \revA{provide} promising opportunities for the identification of neutrino emitters \citep[e.g.,][]{Murase18}. 
Because of their highly variable nature, blazars can undergo strong flaring episodes during which an accompanying IceCube neutrino signal would stand out from the background. Besides, pre-selecting the time window of a blazar flare also mitigates the look-elsewhere effect associated with a blind point-source search in the entire neutrino data set \revB{\citep[see e.g.,][]{IceCube2021},} further improving the sensitivity of these searches.

The first evidence for \revHESSc{a candidate TeV-PeV} extragalactic neutrino source involved the $\gamma$-ray blazar TXS~0506+056. The coincident detection of the neutrino event, \revHESSc{IceCube-170922A}, \revB{and the temporally correlated $\gamma$-ray flaring activity from TXS 0506+056 in 2017 \citep{IceCube2018MMA}, combined with an excess of $13\pm5$ muon-neutrino events observed between 2014 and 2015 by IceCube \citep{IceCube2018}, suggested a possible association of the neutrino emission with the blazar. The lack of \g{} counterpart during the neutrino flare in 2014--2015 and the low \g{} flux seen by \fermi{}-LAT from NGC 1068 compared to the neutrino flux \citep{IceCube2022_NGC1068}, however, suggest that \gs{} may be absorbed in a dense radiation field that acts as an efficient target for neutrino production through photohadronic processes. Therefore, it is still unclear whether or not astrophysical neutrinos are associated with blazars and blazar flares. Further multi-messenger data on other candidate sources are important to understand the conditions and mechanisms for potential neutrino emission from blazars. }

A recent opportunity to explore the connection between neutrinos and high-energy blazars came with the announcement of the neutrino event \neu{} \citep{2021GCN.31191....1I} detected by IceCube as a track-like event with an energy $E_\nu\approx 171$ TeV \revB{and a 50.2\% probability of being astrophysical origin} \footnote{\url{https://gcn.gsfc.nasa.gov/notices_amon_g_b/136015_21306805.amon}} on December 8, 2021. The \revA{gamma-ray} blazar PKS 0735+178 (redshift $z=0.45$) \revHESS{is located} immediately outside of the 90\% error region (2.13$^\circ$; statistical error only) for the neutrino event, 2.2$^\circ$ away from the best-fit position.
Additionally, the Baikal-GVD experiment detected a high-energy neutrino candidate event with an energy $E_\nu\approx 43$ TeV \citep{2021ATel15112....1D} approximately four hours after the IceCube event, about 
\revB{$4.7^{\circ}$ (with an estimated 68\% containment point spread function of $8.1^{\circ}$) from the position of PKS~0735+170.} 
KM3NeT neutrino detectors found one up-going muon neutrino candidate ($\sim$18 TeV) on December 15 in \revB{spatial} coincidence with PKS~0735+178 with a $p$-value of 0.14 \citep{2022ATel15290....1F}. The Baksan Underground Scintillation Telescope reported the observation of a GeV neutrino candidate event four days before \neu{} \citep{2021ATel15143....1P}. Multiwavelength observations of PKS~0735+178 \revB{revealed} flaring states in the radio band \citep{2021ATel15105....1K}, optical band \citep{2021ATel15098....1Z}, X-ray band \citep{2021ATel15102....1S,2021ATel15109....1D,2021ATel15113....1F}, and GeV \g{} band \citep{2021ATel15099....1G}, \mpon{which are potentially associated with the neutrino event \neu{}}. To follow up, we triggered \nustar{} \revA{observations} and provided precise measurements on December 11 and 13, 2021 of the X-ray spectrum \mpon{that was found to be harder than that seen with} \swift{}-XRT \citep{2021ATel15113....1F}. In addition, the imaging atmospheric Cherenkov telescopes (IACTs) \revB{the Very Energetic Radiation Imaging Telescope Array System (VERITAS)} and \revHESSc{the High Energy Stereoscopic System (\hess{})} performed target-of-opportunity (ToO) observations that yielded upper limits above \revHESSc{100~GeV}. 

Recently, \citet{Sahakyan:2022nbz} explored the connection between PKS~0735+178 and \neu{} and reported on \fermi{}-LAT, \swift{}, and optical observations of \revHESS{the blazar}. They compared a proton-synchrotron model and two lepto-hadronic models to explain the spectral energy distribution (SED) of the source and found that a lepto-hadronic model with PeV protons (proton luminosity about three times the Eddington luminosity) interacting with an external target photon field yielded the highest neutrino rate (0.067 neutrinos in three weeks). \revB{A similar study including \nustar{} data was carried out by \citet{Prince23}, who argued through numerical modeling that neutrino production is insufficient without external photons, assuming the neutrino event and the photons are associated.}

In this work, we report on multiwavelength observations of the blazar PKS~0735+178, including those from \nustar{}, VERITAS, and \hess{}, contemporaneous with the IceCube \revB{astrophysical} neutrino candidate \neu{}. 
Since the blazar is outside of the 90\% error region of the IceCube event by $\sim0.1^\circ$, and there are a large number of \g{} blazars that exhibit strong flares, \revB{it is uncertain whether or not there is an association between the flaring events across the electromagnetic spectrum and the neutrino event.} 
We interpret the broadband SED of the source in the context of both leptonic and lepto-hadronic models, and discuss whether the neutrino event could originate from the blazar.

\section{Observations and Data Analysis}
\label{sec:obs_data}
%
\subsection{VERITAS}
%
VERITAS is an array of four IACTs located in southern Arizona  \citep[\revHESSc{31$^\circ$40'30"N 110$^\circ$57'07"W}, 1.3 km a.s.l.;][]{Park15}. It is capable of detecting \gs{} with energies from 85 GeV to $>$30 TeV, with an energy resolution of $\sim$15\% (at 1~TeV) and a \revB{point spread function} of $\sim$0.1$^{\circ}$ (68\% containment at 1~TeV). A point source of 1\% the Crab Nebula flux is detectable by VERITAS at a statistical significance of five standard deviations (5 $\sigma$) in $\sim$25~hours. 

\revA{Independently of} the IceCube trigger, the blazar PKS~0735+178 was \revA{previously} observed by VERITAS between December 2010 and February 2011 for 5.2 hours. The source was not detected and a differential flux upper limit \revB{at 99\% confidence level (C. L.)} \revHESS{of} $9 \times10^{-12}\;\text{cm}^{-2}\; \text{s}^{-1}\; \text{TeV}^{-1}$ at 260 GeV \mpon{was derived} \citep{Archambault16}.

\revA{In response to} the IceCube and multiwavelength alerts, VERITAS collected about 20 hours of quality-selected data on PKS~0735+178 between December 9, 2021, and January 8, 2022, at an average zenith angle of 20.2$^{\circ}$. Observations were performed using the standard “wobble” observation mode \citep{Fomin1994} with a 0.5$^{\circ}$ offset. \revB{The source was not significantly detected}. 
A statistical significance of 3.4 $\sigma$ \revB{on the excess was derived using events} within a 0.1$^\circ$ region around the source.
The integral flux upper limit above 220 GeV, which is the highest energy threshold among these observations, is $2.55\times10^{-12}\;\text{cm}^{-2}\; \text{s}^{-1}$ at 99\% C. L. assuming a power-law spectrum with a photon index of \revHESS{3} \citep[following][]{Rolke05}. 

The \revB{new} VERITAS data were analyzed using the software described in \citet[][]{Cogan08}, with a shower-image template maximum-likelihood reconstruction method and cuts optimized for lower-energy \g{} events \citep[see e.g.][]{Christiansen2017}, and independently confirmed with another analysis software described in \citet[][]{Maier17}. Two nearby \revHESSc{\fermi{}-LAT} sources 4FGL~J0738.4+1539 and 4FGL~J0743.1+1713 were excluded \revB{from the background estimation} with an exclusion radius of 0.3$^\circ$.

\subsection{\hess{}}
Located in the Khomas Highland of Namibia ($23^{\circ}16'18''$ South, $16^{\circ}30'00''$ East), at an elevation of 1800~m above sea level, \hess{} is the only IACT array in the Southern Hemisphere \revHESSc{\citep[][]{hess_aharonian}}.  
\revHESSc{It consists of four 12-m telescopes (CT1-4) placed \revHESS{in a square} with a side length of 120~m, a field-of-view (FoV) of $5^{\circ}$. A fifth 28~m telescope (CT5) \citep[][]{hess_bolmont_2014} was added in 2012 at the center, but it is not used in this analysis.}

Since 2012, \hess{} has conducted a Neutrino-ToO program searching for spatial and temporal correlations between neutrinos and very-high-energy $\gamma$-ray emission. Triggered by the ToO alert from IceCube on December 8, 2021, \hess{} observed in the direction of \neu{} for a total of $16~\mathrm{h}$ from December 8 to 15, 2021, \revHESSc{at an average zenith angle of $42.2^\circ$}.
Only 3.8~h of data were selected based on strict criteria on weather conditions and instrumental status. Observations were performed in wobble mode \revHESS{at an offset from the center of the camera of $0.5^\circ$} \citep[][]{hess_aharonian_2006}. 

The data were analyzed using the \mpon{method} described in \citet[][]{hess_denaurois_2009} with standard gamma-hadron separation and event selection cuts. A circular region of interest (RoI) of 0.1$^\circ$ centered on the position of PKS~0735+178 was defined, and two regions of 0.25$^\circ$ radius around  the nearby sources 4FGL~J0738.4+1539 and 4FGL~J0743.1+1713 were excluded \revB{from the background estimation}. The background was determined using the standard ``\revHESS{reflected} background" technique \citep[][]{hess_berge_2007}. 
The results were validated by a second analysis which uses an independent event calibration and reconstruction \citep[][]{hess_parsons_2014}.

No significant \g{} excess above the expected background \qf{was} detected from the direction of PKS~0735+178. 
The integral flux upper limit \qf{at 99\% C. L.} in the energy range between 0.1~TeV and 10~TeV is $1.82~\times~10^{-11}$~ cm$^{-2}$~s$^{-1}$ assuming 
a power-law spectrum with a photon index of \revHESS{3} \cite[following][]{hess_rolke_2005}. 
For the upper limit calculations, the minimum energy was chosen \revHESS{as the energy where the effective area reaches $10\%$ of its maximum value}, while the maximum energy was chosen such that the number of background events $N_{\mathrm{OFF}}\geq 10$.

\subsection{{\it Fermi}-LAT}
\label{subsec:LAT}
The Large Area Telescope (LAT) onboard the \revHESSc{\fermi{}-LAT} satellite is a pair-conversion \g{} telescope sensitive to \gs{} with energies from $\sim$20~MeV to $>$300~GeV from a $>2$ sr FoV \citep{Atwood09}. 

A binned likelihood analysis was performed for the {\it Fermi}-LAT data using the \textit{Fermi} Science Tools version 11-05-03 \footnote[1]{\url{http://fermi.gsfc.nasa.gov/ssc/data/analysis/software}} 
and \textit{FermiPy} version 1.0.1~\footnote[2]{\url{http://fermipy.readthedocs.io}} 
 \citep{wood2017fermipy} in conjunction with the latest \textit{PASS} 8 instrument response functions \citep{Atwood13}.
Photon events were selected from an energy range between 100 MeV and 300 GeV and a RoI of radius 15$^{\circ}$ centered on the location of PKS~0735+178, within a maximum zenith angle of \revB{90}$^{\circ}$. 
A spatial binning of 0.1$^{\circ}$~pixel$^{-1}$ and four \revB{logarithmic} energy bins per decade were used.
The initial model consisted of all sources within 20$^{\circ}$ of the center of the RoI, based on the spatial positions obtained from the 4FGL Data Release 2 \citep[4FGL-DR2;][]{4FGL} catalog, 
and the templates for isotropic and Galactic diffuse emission, iso\_P8R3\_SOURCE\_V2\_v1.txt and gll\_iem\_v07.fits, respectively. 
\revHESS{An i}terative optimization of the model was performed, removing weak sources with a TS $<$ 10 
and searching for any additional point sources (TS $\geq$ 10) not accounted for in the 4FGL-DR2 catalog 
at each iteration. 

PKS~0735+178 is included in the 4FGL catalog as 4FGL J0738.1+1742, with a peak energy $E_\textrm{peak}\approx(2.1\pm0.3)$ GeV \revA{in the $\nu F_\nu$ representation estimated from} the preferred log-parabola model $dN/dE = N_0 \left( E/E_\textrm{b} \right)^{-\alpha-\beta\log(E/E_\textrm{b})}$, where $E_\textrm{b}$ is the energy scale, 
$E$ is the energy, $\alpha+\beta\log(E/E_\textrm{b})$ is the energy-dependent photon index, and $N_0$ is the normalization. \revA{In this work,} the \fermi{}-LAT spectrum of the source \revA{between December 1 and 28, 2021} was fit with a log-parabola model with $E_\textrm{b}$ fixed at 1.54 GeV. 
The best-fit parameters are $N_0 = (1.95 \pm 0.16) \times10^{-11}\; \text{MeV}^{-1}\;\text{cm}^{-2}\; \text{s}^{-1}$, $\alpha = 2.03\pm0.05$, and $\beta = 0.04\pm0.03$, \mpon{yielding the} best-fit peak energy as $E_\textrm{peak}\approx 0.7$ GeV and its 1-$\sigma$ C. L. interval as roughly 200 MeV -- 2.5 GeV. \revB{The uncertainties are statistical only. Monthly and daily binned light curves between 100 MeV and 300 GeV were made from likelihood analyses freezing the spectral curvature parameters $\alpha$ and $\beta$ while leaving the normalization free.} 

\revD{To check for GeV spectral variability, we conducted a likelihood analysis with daily bins between December 1 and 28, 2021, leaving $\alpha$ and $\beta$ free. The best-fit values for $\alpha$ and $\beta$ are consistent with being constant with probabilities ($p$-values) of 0.55 and 0.88, respectively. The \mpo{curvature parameter, $\beta$, is poorly constrained in the daily spectra} due to limited statistics \mpo{and is} consistent with 0. On Dec 10 and 13, 2021, the best-fit $\alpha$ values are $2.30\pm0.32$ and $2.07\pm0.19$, respectively, in agreement with the average value $2.03\pm0.05$ used for modeling.}

\subsection{\nustar{}}
\label{subsec:nustar}
\nustar{} is a hard X-ray space telescope consisting of two co-aligned optics and two focal plane detectors (FPMA and FPMB) covering a $13' \times 13'$ FoV. \nustar{} is sensitive to 3 - 79 keV photons with an on-axis point spread function of 18" \citep[FWHM;][]{Harrison_2013}.

\revHESS{Two \nustar{} ToO observations of PKS~0735+178 were triggered by the \neu{} alert, in combination with the flux increase from the PKS~0735+178 detected by \swift{}-XRT and \fermi{}-LAT.}
Observation 1 (ObsID = 80701621002, exposure = 22 ks) was performed on December 11, 2021, and observation 2 (ObsID = 80701621004, exposure = 22 ks) was performed on December 13, 2021. The two observations were separated by two days to test the \revA{flux and spectral} variability of PKS~0735+178.

The \nustar{} data were processed using \nustar{} data analysis software (\texttt{NuSTARDAS}) version 2.1.1 contained within \texttt{HEASOFT} version 6.29 along with the \nustar{} calibration database (\texttt{CALDB}) version 20211202. The source and background spectra were extracted from a circular region of radius $30"$ and a box region of $2' \times 2'$, respectively. The source spectra were binned such that each bin has minimum 50 counts. Using \texttt{XSPEC} \citep{Arnaud96}, the binned spectra from FPMA and FPMB of each observation were simultaneously fit between 3 and 40 keV \revHESS{beyond which} background begins to dominate. Observation 1 yielded 493/517 (net/total) counts for FPMA and 469/521 counts for FPMB. Observation 2 yielded 386/412 counts for FPMA and 395/433 counts for FPMB.

The spectra were well fit by a single power law with a constant factor (cross-normalization) to account for the difference between the two detectors (\texttt{cons*po}). Adding an absorption component (\texttt{cons*tbabs*po}) does not improve the fit quality or constrain the hydrogen column density. The total neutral hydrogen column density is low in the direction of the source \citep[$4.48 \times 10^{20} \textrm{ atoms cm}^{-2}$;][]{Willingale13} \footnote[3]{\url{https://www.swift.ac.uk/analysis/nhtot/index.php}}, thus the absorption in the \nustar{} hard X-ray band is negligible. 
The best-fit photon index $\Gamma$ of observation 1 is \qf{$1.85 \pm 0.06$} (1 $\sigma$ statistical error) with a reduced $\chi ^2=1.08$ for 31 degrees of freedom (dof). 
Observation 2 exhibits slightly harder \revHESS{spectrum} than observation 1, yielding the best-fit $\Gamma=1.70 \pm 0.07$ with a reduced $\chi ^2=1.06$ for 26 dof. 

\subsection{{\it Swift}-XRT}
\label{subsec:XRT}
The X-Ray Telescope (XRT) on the Neil Gehrels {\it Swift} Observatory, sensitive to energies from $\sim$0.2~keV to 10~keV \citep{Gehrels04, Burrows05}, observed PKS~0735+178 nine times in photon counting mode for a total exposure of $\sim$15.2 ks from December 10, 2021 to January 6, 2022 in response to the IceCube alert. Prior to this event, 12 {\it Swift} observations were taken between December 2009 \revHESSc{and} October 2011, with a total XRT exposure of $\sim$21.3 ks. The \swift{}-XRT light curve was retrieved from the public online tool ``the \swift{}-XRT data products generator"~\footnote[4]{\url{https://www.swift.ac.uk/user_objects/index.php}} \citep{Evans07, Evans09} and is shown in Figure~\ref{fig:lc}. The \swift{}-XRT spectral analysis was performed on data taken on December 10 and 13, 2021 (ObsID 00036372014 and 00036372016, respectively) using \texttt{HEASOFT} version 6.29 and (\texttt{CALDB}) version 20210915. The source and background spectra were extracted from a circular region of radius 20 pixels and an annulus region with inner and outer radii of 70 and 120 pixels, respectively. The source spectra for the observation on December 10 and 13 were grouped requiring a minimum of 20 and 10 counts per bin, respectively. An absorbed power law with the neutral hydrogen column density frozen \revHESS{at} $4.48 \times 10^{20} \textrm{ atoms cm}^{-2}$ \revHESS{was} used to fit the spectra, yielding photon indices of $2.71\pm0.09$ ($\chi ^2 / \textrm{dof} = 11.8/18$) and $2.68\pm0.16$ ($\chi ^2 / \textrm{dof} = 18.3/15$), respectively. Note in Figure~\ref{fig:sed} that the \swift{}-XRT spectrum on December 13 captures the \revHESSc{low-energy end of the high-energy SED component} above a few keV, consistent with the \nustar{} measurements on the same night. 

\subsection{{\it Swift}-UVOT}
\label{subsec:UVOT}
The {\it Swift}-Ultraviolet/Optical Telescope (UVOT; \citealt{2005SSRv..120...95R}) is capable of detecting optical to UV photons with six filters with central wavelengths of $V$ 5468 \AA, $B$ 4392 \AA, $U$ 3465 \AA, $UVW1$ 2600 \AA, $UVM2$ 2246 \AA, and $UVW2$ 1928 \AA. The {\it Swift}-UVOT observations of PKS~0735+178 were analyzed using \texttt{HEASOFT} version 6.29 and (\texttt{CALDB}) version 20211108. A source region with a radius of $5.0''$ and a background region of the same size were used to extract signal and background counts. The magnitude of the source was then computed using \texttt{uvotsource} and converted to flux using the zero-point for each of the UVOT filters from \citet{Poole08}. 
\revHESS{The e}xtinction correction was applied following \citet{2009ApJ...690..163R}, using the color excess E(B-V) = 0.0292 \citep{SandF2011}. 

\subsection{Other Multiwavelength Facilities}



PKS~0735+178 was monitored by many optical facilities. 
For this study, we used the optical magnitudes of PKS~0735+178 in the $g$ band from the publicly available \qfn{aperture photometry results from} the All-Sky Automated Survey for Supernovae \citep[ASAS-SN;][]{2014Shappee, 2017Kochanek}~\footnote[5]{\url{https://asas-sn.osu.edu/}}, in the $R$ band from the Asteroid Terrestrial-impact Last Alert System \citep[ATLAS;][]{Tonry18, Heinze18}, \revB{in the $B$ and $R$ bands from the Automatic Telescope for Optical Monitoring \footnote{\url{https://www.lsw.uni-heidelberg.de/projects/hess/ATOM/}} \citep[ATOM;][]{ATOM}}, and in the $K_s$ band reported on the Astronomer's Telegram from the Nordic Optical Telescope \citep{2021ATel15136....1L}. 
We also used the 37 GHz results from the Owens Valley Radio Observatory \citep{2021ATel15105....1K}. The ASAS-SN, ATLAS, \revB{and ATOM} light curves are shown in Figure~\ref{fig:lc}, and the fluxes in all bands mentioned above \revB{measured between one and eight days after} \neu{} are shown in the SEDs in Figures~\ref{fig:sed} and \ref{fig:sed_hadro}. \revB{Archival data from radio to UV band were taken from the ASDC SED Builder Tool of Italian Space Agency \citep{Stratta2011} including data from various catalogs and databases \citep{Planck2011A&A...536A...7P, 1981A&AS...45..367K, 2009ApJS..180..283W, 2010AJ....140.1868W, 2012A&A...541A.160G, VizieR2000A&AS..143...23O}. }


\section{Results}
\label{sec:res}
%
\subsection{Temporal Variability}
\label{sec:res_var}

\begin{figure*}[hb]
\centering
\subfigure[The long-term variability of PKS~0735+178. The flux above 100 MeV measured by \fermi{}-LAT starting in 2008 (\textit{top}) is binned by month, the count rate between 0.3 and 10 keV measured by \swift{}-XRT (\textit{middle}) and the optical $B$, $g$, $V$, and $R$ band magnitudes from ASAS-SN, ATLAS, and ATOM (\textit{bottom}) are binned by observations. The dashed lines show the time of the neutrino event \neu{}. ]{
    \includegraphics[width=0.48\textwidth]{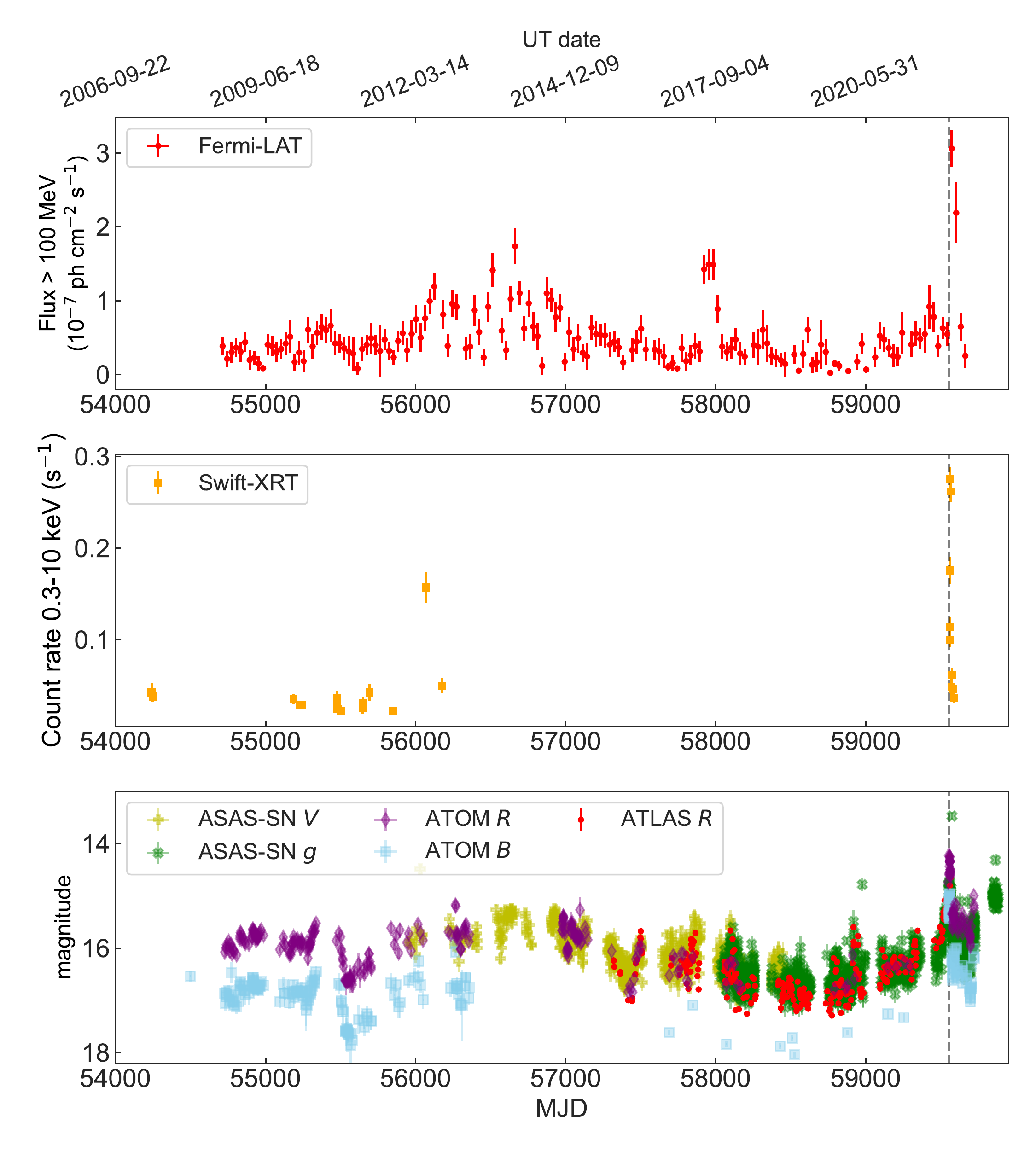}
    \label{fig:lc1}
}
\subfigure[The light curves focusing on a $\sim$2-month period after the neutrino event. The \fermi{}-LAT flux is binned nightly. The vertical dashed lines show the time of the neutrino event; the dotted lines show the times of the two \nustar{} observations presented in this work; the green hatch-filled regions show the period during which the VERITAS and \hess{} data were taken. \revA{The dashed curve in the middle panel shows the best-fit model with three exponential components to describe the \swift{}-XRT data.}]{
    \includegraphics[width=0.48\textwidth]{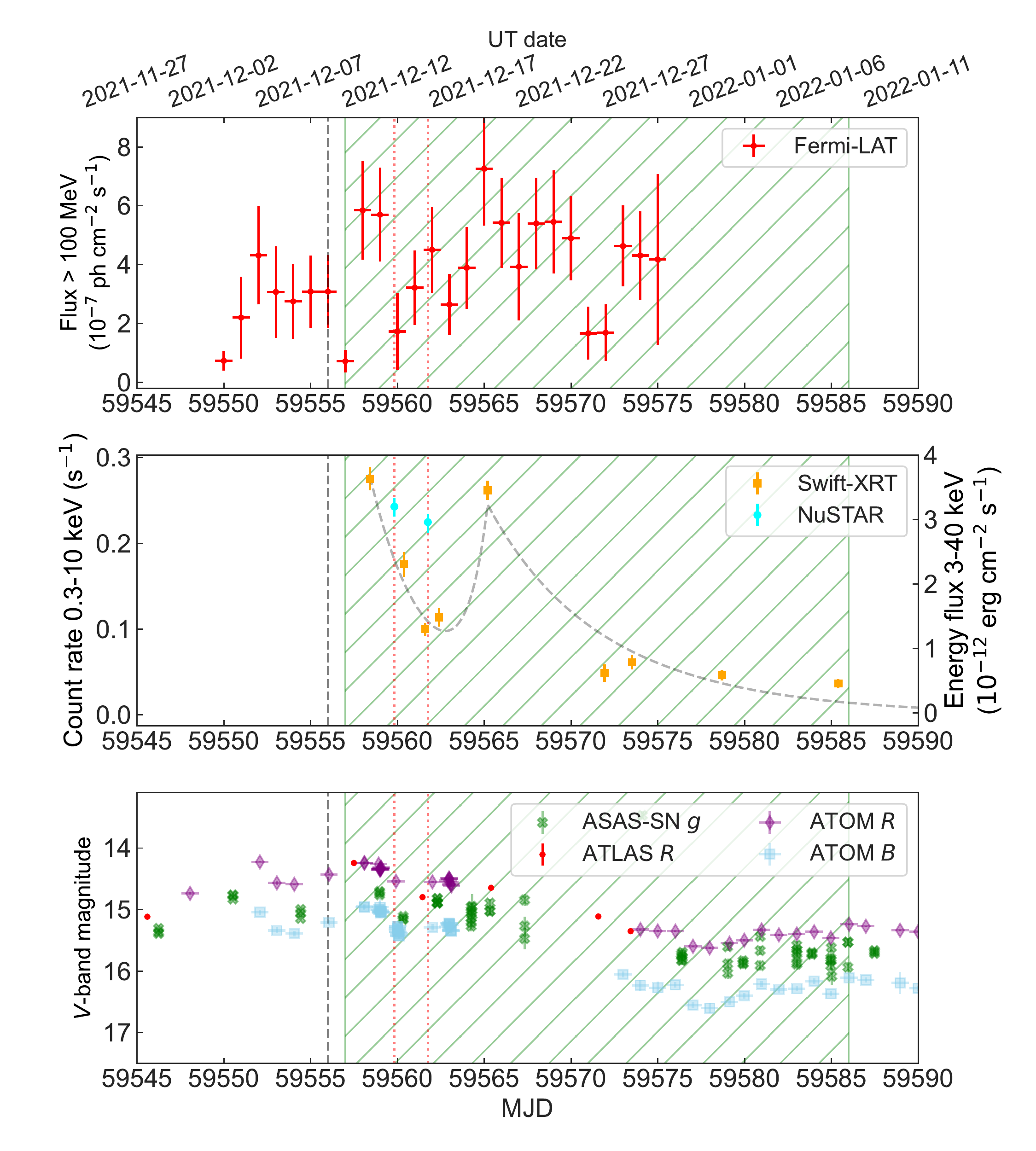}
    \label{fig:lczoom}
}
\caption{The light curves of PKS~0735+178 measured by \fermi{}-LAT, \nustar{}, \swift{}-XRT, ASAS-SN, ATLAS, and ATOM.  }
\label{fig:lc}
\end{figure*}

The monthly \g{} and the observation-wise X-ray \qfn{and optical} light curves \mpon{are} shown in Figure~\ref{fig:lc1} \mpon{and} illustrate that around the time of the neutrino event \neu{}, the blazar \src{} exhibited the highest fluxes in \qfn{all these data sets}. 
The flux above 100 MeV reached a peak of $(3.1\pm0.3)\times 10^{-7}\;\text{photon} \;\text{cm}^{-2} \;\text{s}^{-1}$ in the monthly bin centered at \revB{December} 24, 2021, roughly seven times higher than the average flux in the 4FGL catalog and twice as bright as the previously highest state in 2014.

Figure~\ref{fig:lczoom} \revHESS{shows} the daily \g{} \mpon{and} observation-wise X-ray \qfn{and optical} light curves during a $\sim$2-month period after the IceCube event, focusing on the highest flux state of the source. 
The soft X-ray flux exhibited daily variability, starting at the highest count rate of $0.28\pm0.01 \;\text{s}^{-1}$ on December 10, 2021, followed by a rapid decay down to about $0.1 \;\text{s}^{-1}$ on December 12 and 13 (with a flux halving time of $2.2\pm0.4 \;\text{d}$), and a rapid rise to the second highest count rate of $0.26\pm0.01 \;\text{s}^{-1}$ on December 17 (with a flux doubling time of $0.8\pm0.4 \;\text{d}$, \revD{which was used to constrain the size of the emitting region in the discussions}). A model with three exponential components (two decaying and one increasing) \revHESSc{was} used to \mpon{characterize} the soft X-ray flux variability \revA{(see Figure~\ref{fig:lczoom})}. \revB{The peak times were fixed to the two XRT observations with the highest count rate, and five parameters, two peak times and three variability timescales, were left free.}
\qfn{The optical observations also revealed a decay in the brightness of the blazar by $\sim$ 1 magnitude within three weeks after \neu{}, while exhibiting daily variability with a smaller amplitude.}

Daily variability above 100 MeV \revB{was present} during this highest-flux period, \revB{with a constant-flux model yielding a poor fit with $\chi^2=79$ for 25 degrees of freedom, which is consistent with the} variability above 300 MeV reported by \citet{Sahakyan:2022nbz}. 
\revD{The strongest evidence for fast GeV \g{} variability comes from the low flux measured on MJD 59557 and the immediately following high fluxes measured on MJD 59558 and 59559. The variability timescale, however, is model-dependent and not well constrained given the large uncertainty in the flux measurement. A fit to a model with three flux peaks yielded the shortest best-fit variability time of $0.38 \pm 0.46$ day, consistent with the X-ray variability mentioned above and the value of 0.35 day reported by \citet[][]{Prince23}. No evidence for short-term spectral variability was found in the GeV spectrum (see Section~\ref{subsec:LAT}). }

No hard X-ray variability was observed between the two \nustar{} observations on December 11 and 13, the 3 -- 40 keV fluxes of which are $(3.2 \pm 0.2)$ and $(3.0 \pm 0.2)$ \mpon{in units of} $10^{-12} \;\text{erg} \;\text{cm}^{-2} \;\text{s}^{-1}$, respectively. 

\subsection{Broadband SED}
\label{subsec:SED}

\begin{figure*}[hb]
\plotone{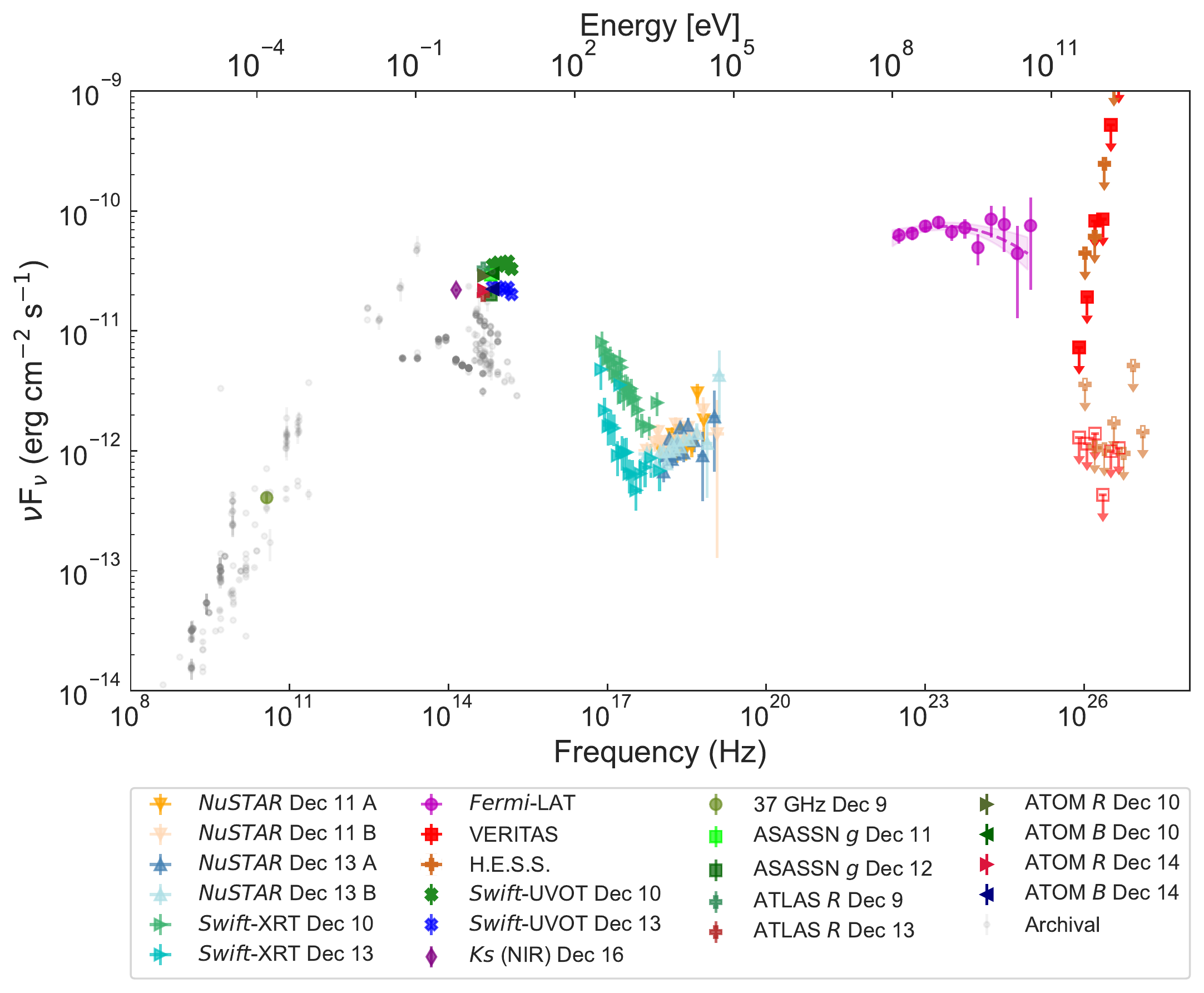}
\caption{The broadband SED of PKS~0735+178 in December 2021 along with archival data from radio to UV bands (shown as gray filled circles). The VERITAS and \hess{} spectra were averaged over the entire period of the ToO observations (see Section \ref{sec:obs_data}), with the observed SED and those corrected for the EBL absorption shown as the \revB{unfilled and filled} symbols, respectively. The \fermi{}-LAT spectrum was averaged over a four-week period from December 1 to 28, 2021. \revB{The magenta shaded region shows the best-fit log-parabola model to the \fermi{}-LAT spectrum and its 1-$\sigma$ uncertainty.}}
\label{fig:sed}
\end{figure*}

The broadband SED of PKS~0735+178 \revB{contemporaneous with} the IceCube neutrino event is shown in Figure~\ref{fig:sed}. \qfn{The SED includes \fermi{}-LAT, VERITAS, and \hess{} data averaged over $\sim$1 month after \neu, and data from X-ray down to optical bands for each observation on a given night \revB{between December 9 and 16, 2021}. Among these observations, an approximately simultaneous SED can be constructed using data from \nustar{}, \swift{}, and ATLAS taken on December 13, 2021. Given the daily variability, as discussed in the previous section, the SED on December 13 should be the focus of the numerical modeling (discussed in Section~\ref{subsec:numericalSED}).} 

The synchrotron and the high-energy peak frequencies are loosely constrained by the rather flat optical/UV and GeV \g{} spectra. Nominally, the highest energy flux is observed at a few times $10^{14}$~Hz and \revHESSc{at $\sim10^{24}$~Hz}, respectively, but the spectral cut-off occurs at a somewhat higher frequency, in particular for the \g{} component. 
The measured soft and hard X-ray spectra fully constrain the tail of the synchrotron emission and the beginning of the high-energy component of the SED, respectively. The transition occurs at a few keV. 
To be noted from the SED is the steep decline of the flux near $100$~GeV. The VERITAS upper limit at $330$~GeV is \revD{particularly relevant. After correction for absorption by extragalactic background light (EBL) \citep[shown as the darker red squares in Figure~\ref{fig:sed};][]{Dominguez11}, the flux limit is} about a factor of ten \revHESSc{below the log-parabola extrapolation of the \fermi{}-LAT spectrum}. There are at least three ways to explain this finding. 
\begin{enumerate}
\item There could be an intrinsic cut-off at about 100 GeV in the observed photon energy. 
\item The redshift $z=0.45\pm0.06$ \citep{2012A&A...547A...1N} could be an underestimate, 
and the true value might be $z\approx 0.8$.
\revHESSc{This would bring the upper limit derived by VERITAS, after correcting for EBL absorption, in line with the power-law extrapolation by \fermi{}-LAT.} 
The redshift of the source was derived assuming that BL Lac host galaxies can be used as standard candles. A lower limit of 0.424 was derived from a Mg II absorption-line doublet \revHESSc{\citep[][]{redshift_lowerlimit}}. The identity of the absorber towards PKS~0735+178 is yet to be determined.
\item There \revHESS{could be} significant \g{} absorption within the broad-line or narrow-line region, whichever exists, which would require \g{} production within a parsec or so from the central engine.
\end{enumerate}
In section~\ref{sec:discuss} we shall discuss the observed SED in the context of the neutrino association.

\section{Discussion}\label{sec:discuss}

\subsection{Model Basics and SSC}\label{sec:4-1}
We shall first discuss the implications of an intrinsic \g{} cut-off at 100 GeV, assuming a one-zone \revHESSc{synchrotron-self-Compton (SSC)} scenario. The synchrotron spectrum \mponn{seems to peak} at a few $10^{14}$~Hz in frequency, or $\epsilon_\mathrm{peak}=\epsilon_\mathrm{peak,eV}$~eV in photon energy, where $\epsilon_\mathrm{peak,eV}$ is close to 1. 
\revA{The synchrotron peak is quite broad and loosely constrained by the rather flat optical to UV spectrum, with UVOT data suggesting an optical/near-UV flux that is a slowly declining function of frequency. Beyond the near-UV band, there are no measurements until the soft X-ray band.} The \swift{}-XRT flux at $300$~eV is a factor of 30 below the UV flux at $3$~eV, indicating that the cut-off energy is likely below $100$~eV, or $\epsilon_\mathrm{peak,eV}\lesssim 100$, and likely far lower based on the slowly declining UV spectrum. 

There \revB{has} \revC{to} be \revHESSc{a SSC contribution} to the SED. If the Thomson limit applies, the ratio between the inverse-Compton (IC) and the synchrotron peak frequencies $(100\ \mathrm{GeV}/\epsilon_\mathrm{peak}$) gives the square of the peak Lorentz factor of electrons, leading to 
\begin{equation}
\gamma_\mathrm{peak}\approx \frac{3\times 10^5}{\sqrt{\epsilon_\mathrm{peak,eV}}}\ .
\label{eq:1}
\end{equation}
The Thomson limit at \revA{$\epsilon_\gamma = 100$~GeV requires $\gamma \gg \epsilon_\gamma/(Dm_ec^2) \approx (8\times 10^3)/D_{25}$}, where $D=25\,D_{25}$ is the Doppler factor, which renders the Thomson limit applicable for a synchrotron peak up to the hard-UV band. 
If the true synchrotron peak energy were close to or beyond this limit, \revA{which is unlikely given the UVOT and XRT observations}, the Klein-Nishina transition would add to the cut-off at $100\ \mathrm{GeV}$.  

In the following, any quantity measured in the jet frame is denoted with a prime (e.g., $R'$). \revC{For the nominal redshift (0.45),} the synchrotron peak frequency and the peak Lorentz factor of electrons constrain the magnetic-field strength in the emission zone,
\begin{equation}
B' D\approx ( 10^{-3}\ \mathrm{G})\,\epsilon_\mathrm{peak,eV}^2\ .
\label{eq:BD}
\end{equation}
Eq.~\ref{eq:BD} indicates that the magnetic field would have to be very weak, unless the true synchrotron peak frequency were large, $\epsilon_\mathrm{peak,eV} \gg 1$. \mpon{Alternatively, the \g{} cut-off may be at a few GeV, but that would leave the observed flux at 10 GeV -- 100 GeV unexplained, and so a weak magnetic field is needed in the one-zone SSC scenario.} 
At the same time, the jet-frame photon energy density at the synchrotron peak would have to be a factor of two or three larger than that in the magnetic field to reproduce the $\nu F_\nu$ flux ratio of the IC and the synchrotron peak. We shall now discuss whether or not that is possible.

Assuming isotropic emission in the jet frame, at the distance $d\approx 8\times 10^{27}\ \mathrm{cm}$ \mponn{\citep[see \revHESSc{Eq.~28} in][]{planck2020}} \revB{from the observer}, the intrinsic 
photon density per logarithmic energy interval and the energy density in the radiation field are
\begin{equation}
U'=\frac{3\,d^2}{c\,R'^2\,D^4}\, \nu F_\nu =\epsilon' \revC{n_{\ln\epsilon}'}
\label{eq:n_eps}
\end{equation}
Because of the flat observed optical spectrum in the $\nu F_\mu$ representation, the energy density of the synchrotron peak should be roughly similar to that at $\epsilon\approx 1$~eV, corresponding to $\epsilon' \approx D^{-1}$~eV. We find 
\begin{equation}
U'\approx
\frac{5\times 10^{-3}\ \mathrm{erg\,cm^{-3}}}
{D_{25}^4\,R_{16}^2}\, ,
\label{eq:u_eps}
\end{equation}
where we denote the radius of the emission zone as $R' = R_{16}\,(10^{16}\ \mathrm{cm})$.

The energy density in the magnetic field (cf. equation~\ref{eq:BD}) required for the SSC \revC{peak to be close to} 100 GeV and the synchrotron \revC{turnover} at $\epsilon_\mathrm{peak,eV}$ is 
\begin{equation}
U_\mathrm{B}'\approx
(6\times 10^{-11}\ \mathrm{erg\,cm^{-3}})\,\frac{\epsilon_\mathrm{peak,eV}^4}
{D_{25}^2}\ ,
\label{eq:u_B}
\end{equation}
which agrees with the energy density of the photon field at the synchrotron peak only if \revC{the synchrotron cut-off energy were high, $\epsilon_\mathrm{peak,eV}\gtrsim 10$, and likewise the size of the} emission zone, $R_\mathrm{16}\gtrsim 10$. More precisely, \revD{our new TeV-band data} require that 
\begin{equation}
\epsilon_\mathrm{peak,eV}\approx \frac{100}{\sqrt{D_{25}\,R_{16}}}\ ,
\label{eq:flux}
\end{equation}
otherwise the high-energy component of the SSC scenario will roll over \mpon{far below} $100$~GeV. As the UVOT data suggest $\epsilon_\mathrm{peak,eV}\approx 1$, and the source radius cannot be made arbitrarily large while keeping SSC dominant, the SSC model has difficulty reproducing a 100-GeV cut-off in the \g{} spectrum. 

The \revB{timescale of the flux variability} at $\epsilon_\mathrm{peak,eV}$ should be at least as large as the total electron cooling time in the observer frame, here about a third of the synchrotron cooling time, and be commensurate with the observed daily flux variations (see Section~\ref{sec:res_var}). We find in the observer frame 
\begin{equation}
t_\mathrm{loss} \gtrsim \frac{1}{3}\,t_\mathrm{syn}
\approx
(0.7\ \mathrm{hours})\,D_{25}^{11/4}\,R_{16}^{7/4} \ ,
\end{equation}
where in the last step we used \revB{Eqs.~\ref{eq:1}, \ref{eq:BD}, and} \ref{eq:flux}. Evidently, a very large emission zone will not allow rapid energy losses. \revHESSc{We should add that the flat $\nu F_\nu$ spectrum between the optical and the peak at $\epsilon_\mathrm{peak,eV}$ \revC{is likely explained by cooling, which} requires an active state that is at least as long as the loss time of electrons radiating in the optical band. This time is about ten times longer than that at the best-fitting cut-off frequency (cf. Eq.~\ref{eq:flux}), but still shorter than one day.}
The fastest observed variability timescale is $\sim1$ day at soft X-ray energies, at which the emitting electrons have an even shorter synchrotron cooling time than that at $\epsilon_\mathrm{peak,eV} $.
The causality argument \revC{limits the radius to} $R_{16} \lesssim 5\,D_{25}$ for daily variations. Diffusive escape primarily affects the inverse-Compton component in a SSC model and here is sufficiently slow, even if the diffusion coefficient is a \revC{few hundred times} the Bohm rate.

We conclude that for a synchrotron peak that extends far into the UV band, the SED and in particular the cut-off near $100$~GeV \revHESSc{could} be reasonably well explained with a one-zone SSC model. \revB{But if} the synchrotron peak indeed lies in the optical, \revB{as the data appear to show}, the high-energy \g{} emission would fall off at significantly lower energy than is observed, $\epsilon_\mathrm{IC,peak}\ll 100\ \mathrm{GeV}$, and the SSC model has difficulty reproducing the observed \g{} spectrum. 

\subsection{Viability of Neutrino Emission} 
\label{subsec:neutrino}
\revHESSc{Since a simple SSC model cannot explain the SED of \revB{the} radiation from PKS~0735+178 unless the true synchrotron peak is hidden in the unobserved far-UV band, we shall now explore the viability of associated neutrino emission, which cannot be accounted for by the SSC scenario alone.} A significant number of accelerated ions or protons \revC{may} also exist in the jet. Neutrino emission through the $pp$ channel requires a \revC{very large} particle density in the jet and hence a large jet power \citep[e.g.][]{2000A&A...354..395P}. More likely, neutrino production occurs through p$\gamma$ interactions \citep{1995PhR...258..173G}. The implications in the case of PKS~0735+178 depend on the origin of the photon field with which the energetic protons would interact \citep{Sahakyan:2022nbz}.

\subsubsection{\revC{Co-spatially} Produced Target Photons}
\label{subsubsec:neu1}
\revHESSc{If there were no absorption \revC{of gamma rays with a few hundred GeV by pair production} \revB{with} ambient photons from the broad-/narrow-line region (BLR/NLR), the low observed flux \revA{upper limit} at 350 GeV would strongly constrain the possibility of neutrino production. This is due to moderate EBL absorption at this energy (for the nominal redshift of $z=0.45$), as well as negligible internal absorption, \revC{unless $D_{25}^5 R_{16}\lesssim 0.03$}.}
The effective area of IceCube for \g{} follow-up Bronze alerts with neutrino energy $E_\nu \approx 170\ $TeV is $A_\mathrm{eff}\approx 30\ \mathrm{m}^2$ \citep{2019ICRC...36.1021B}, and one 170-TeV neutrino event within a month would correspond to a neutrino flux 
\begin{equation}
E F_\nu (E) \approx 1.5\times 10^{-10} \ \mathrm{erg\,cm^{-2}\,s^{-1}}
\label{nuflux}
\end{equation}
that extends over a decade in neutrino energy. The true neutrino flux, however, should be much lower than that \revC{estimate}, otherwise the hard X-ray flux from secondary electrons \mponn{($p\gamma$ pair production)} in PKS~0735+178 would exceed the observed level \citep[e.g.,][]{2019NatAs...3...88G}. 

The \revC{cascade emission of charged and neutral pions in the $100$-TeV band} provides a \g{} flux at a few hundred GeV that is similar to the neutrino flux at $ 170\ $TeV \citep{2017ApJ...843..109G}. The VERITAS upper limits are about a factor 30 lower \revHESS{than} that. If there is no strong \g{} absorption above about $100\ $GeV, then the expected rate of neutrinos would be only 0.03 per month and the association would likely be a chance event.

\revHESSc{The proton energy is about 20 times the energy of the emergent neutrino, assuming interaction through the $\Delta$ resonance \citep[e.g.,][]{2010ApJ...721..630H}. This is justified, a), by the steeply declining number spectrum of target photons between the optical and hard X-rays and, b), as protons of higher energies would \mponn{also produce} neutrinos at higher energy.} In fact, at higher energies the IceCube effective area is even larger, thus leading to a large expected number of observed neutrinos at an energy higher than that of the event in question. The proton Lorentz factor in the jet frame is
\begin{equation}
\gamma_p' \approx \frac{1+z}{D} \frac{20\, E_\nu}{m_p c^2}\approx 2\times 10^5 D_{25}^{-1}\ .
\end{equation}
The typical energy of the target photons in $p\gamma$ interactions is 
\begin{equation}
\epsilon_\mathrm{target}' \approx \revC{\frac{200\,\mathrm{MeV}}{\gamma_p'}}
\approx D_{25}\,(1\ \mathrm{keV}).
\label{eq:epsilon}
\end{equation}
In the observer frame, that would be $25$~keV, i.e. in the energy band covered by \nustar{}. Assuming the hard X-rays are produced co-spatially with the neutrinos and isotropically in the jet frame, we can estimate the jet-frame photon density as a function of the radius of the emission zone, $R'$. In analogy to the derivation of equation~\ref{eq:u_eps}, but for the hard X-rays with flux $\nu F_\nu \approx 2\times 10^{-12}\ \mathrm{erg\,cm}^2\,\mathrm{s}^{-1}$, we find for the jet-frame photon density 
\begin{equation}
n_{\ln\epsilon}'\approx \frac{2\times 10^5\ \mathrm{cm^{-3}}}{D_{25}^4\, R_{16}^2} .
\end{equation}
The cross section of $p\gamma$ interaction is $\sigma_{p\gamma} \approx 5\times 10^{-28}\ \mathrm{cm^2}$. With the inelasticity $K=0.2$ \citep{2000CoPhC.124..290M} we find for the energy-loss time of the radiating protons
\begin{equation}
t_{p\gamma}'\approx\frac{1}{K\,c\,n_{\ln\epsilon}'\,\sigma_{p\gamma}}
\approx D_{25}^4\, R_{16}^2\ \left(1.7\times 10^{12}\ \mathrm{s}\right) .
\end{equation}
In the observer frame, that would correspond to a few thousand years. Protons will diffusively escape from the emission zone in a much shorter time, even for Bohm diffusion, and hence $p\gamma$ interactions are \revC{an inefficient loss process}. To reach any kind of reasonable output, we need a source radius $R_{16}\lesssim 1$. That would also be called for, if the emitter were located inside the BLR/NLR.

The most important unknown input in an estimate of the proton injection luminosity is the observed duration of the active state, $T_\mathrm{act}$. Building on Eq.~\ref{nuflux} we find the neutrino luminosity that gives one neutrino as
\begin{equation}
L_\nu'\approx \frac{6\times 10^{41} \ \mathrm{erg/s}}{D_{25}^4} \,\left(\frac{30\ \mathrm{days}}{T_\mathrm{act}}\right)\, .
\end{equation}
The required proton injection luminosity is at least a factor $t_{p\gamma}'/T_\mathrm{act}'$ larger than the neutrino luminosity, 
\begin{equation}
L_p'\gtrsim  \frac{R_{16}^2}{D_{25}}\,\left(\frac{30\ \mathrm{days}}{T_\mathrm{act}}\right)^2\,\left(1.5\times 10^{46} \ \mathrm{erg/s}\right) .
\end{equation}
In the AGN frame, the kinetic power of the jet is approximately a factor $D^2$ larger \mpon{than that} and would exceed the Eddington limit \citep[see e.g., ][]{2014Natur.515..376G}, which is $L_\text{edd}\approx 10^{47} \;\text{erg}\; \text{s}^{-1}$ for $M_\text{BH}\approx 8\times10^8 M_\odot$ \citep{Ghisellini10}, unless the activity state lasts much longer than 30 days, in fact at least half a year. The same problem arises for TXS~0506+056. Given the high X-ray flux (which could largely be produced by secondary electrons), it is possible to reproduce the SED of \revHESSc{PKS~0735+178} with a lepto-hadronic scenario, if the power requirement could be reduced \citep[see also][]{Sahakyan:2022nbz}. 

\subsubsection{Target Photons from the BLR/NLR}
\revC{External photon fields may potentially reduce the power requirement. For example, spine-sheath models \citep[e.g.][]{2015MNRAS.451.1502T} require less power but imply a coherent bright sheath that is \mpon{at least ten} parsecs long, otherwise the activity cannot \revC{last more than} a few weeks.}

The featureless optical spectrum of PKS~0735+178 suggests that it may be a BL Lac. It has been noted in BL Lacs that the jet may pass through a region harboring a significant jet-external photon field \citep{2022ApJ...926...95F} that for simplicity we \revC{henceforth refer to as} BLR. The BLR is an external source of photons, with which high-energy protons could interact and produce neutrinos. \revC{If BLR emission would be the dominant source of target photons, then} the \revB{particle acceleration site} cannot be comoving with the jet. If it were, it would leave the BLR within \mponn{a few days of observed time, as in that time span the jet travels about a parsec for a Doppler factor $D=25$.} This \revC{decreases} the Doppler boosting of the emission, but relaxes the factor $D^2$ that would otherwise govern the relation between the jet-frame proton injection luminosity and the kinetic power of the jet, and likewise for leptonic models \citep[e.g.][]{2004ApJ...613..725S}.

To estimate the photon density inside the BLR, we assume that the accretion disk produces 10\% of the observed UV flux, about 10\% of which is rescattered. For a BLR of radius $R_\mathrm{BLR}=2$~pc and monoenergetic photons of energy $\epsilon_\mathrm{BLR}\approx 15$~eV, the photon density within the BLR in the AGN frame is 
\begin{equation}
n_{\ln\epsilon}\approx  7\times 10^5\ \mathrm{cm^{-3}} 
\ .
\end{equation}
The pair-production opacity within the BLR would be around unity for photons of about 200~GeV \citep{2010ApJ...717L.118P}, consistent with a scenario in which the sharp drop of observed flux near 100~GeV is at least partially caused by absorption in the BLR.

In the jet frame, the photon density would increase by the jet Lorentz factor, $\Gamma$, and the mean energy of BLR photons would be 
\begin{equation}
n_{\ln\epsilon}' \approx \Gamma_{25}\,(2\times 10^7\ \mathrm{cm^{-3}}),\quad
\epsilon_\mathrm{BLR}'\approx \Gamma_{25}\,(500\ \mathrm{eV})\ ,
\label{eq:16}
\end{equation}
which is reasonably close to that estimated in Eq.~\ref{eq:epsilon} \mpon{and hence permits a rescaling of the estimate for internal photon fields}. 

The energy-loss time of the radiating protons would be
\begin{equation}
t_{p\gamma}'
\approx \frac{1.7\times 10^{10}\ \mathrm{s}}{\Gamma_{25}} \ ,
\end{equation}
and the proton injection luminosity would be
\begin{equation}
L_p'\gtrsim  \frac{1.5\times 10^{44} \ \mathrm{erg/s}}{\Gamma_{25}\,D_{25}^5} \left(\frac{30\ \mathrm{days}}{T_\mathrm{act}}\right)^2\, .
\end{equation}
The jet luminosity is about a factor $D^2$ larger (about $10^{47} \;\text{erg}\; \text{s}^{-1}$) and in marginal agreement with the Eddington luminosity \citep{2002AJ....123.2352X}.

\subsection{External Compton Scattering}
\label{subsec:ec}
The emission zone radius is not constrained in an external hadronic model and can be determined by the leptonic emission processes. IC scattering of BLR photons would be strongly Klein-Nishina-suppressed for electrons with a Lorentz factor exceeding a few hundred. Radiation modeling of blazars often requires a hard electron spectrum, $N(\gamma)\propto \gamma^{-2}$ or similar, \revC{with Lorentz factors up to a few thousands}. The inverse-Compton output would then be a narrow spectral component at the Klein-Nishina transition, $E_\mathrm{KN}\approx 10$~GeV, with a hard spectrum at lower energies, $\nu F_\nu \propto \nu^{0.5}$, that asymptotically transitions to a steep decline well above a few tens of GeV \citep[e.g.,][]{Sikora09}, \revD{which is well in line with the new flux limits in the TeV band that we present}. 
The same pair absorption in the BLR as discussed in Section~\ref{subsec:neutrino} also introduces a spectral cut-off around 200 GeV. The Klein-Nishina effect and the pair-absorption process will both contribute to the steep decline of the observed flux around 100 GeV.
The peak flux of the upscattered BLR photons at about $10$~GeV can be scaled to the observed flux at $1\ $MeV, to which the same electrons would up-scatter the optical photons at the synchrotron peak \revHESSc{in the Thomson regime}. 

For our fiducial parameters $D_{25}$ and $R_{16}$ the $\nu F_\nu$ \mpon{flux} at $10\ $GeV from up-scattering BLR/NLR photons may be a factor of $10$ higher than the SSC flux at $1\ $MeV, and even higher for a larger Doppler factor. It then appears \revC{viable} to produce the observed radiation at $10\ $GeV to $100\ $GeV with external Compton scattering, which permits the SSC component to fall off at a few GeV. The corresponding reduction in $\gamma_\mathrm{peak}$ (cf. Eq.~\ref{eq:1}) permits a stronger magnetic field than estimated in Eq.~\ref{eq:BD}.  Reproducing a comparable $\nu F_\nu$ flux of the synchrotron and the \g{} component in the SSC scenario would no longer require a high value of $\epsilon_\mathrm{peak,eV}$. \revC{This alleviates the requirement of a strong unseen UV synchrotron component that  in an SSC model we found necessary to explain all \g{} data up to $100\ $GeV (see Section~\ref{sec:4-1}).
It also implies that the cut-off near $100$~GeV is partially caused by pair production with BLR photons and partially by the maximum energy of the external-Compton emission component.}

\subsection{\mc{Numerical SED Modeling}}
\label{subsec:numericalSED}
Starting from the analytical estimates discussed in the previous section, we performed a numerical simulation of the photon and neutrino emission from \src\ during the December 2021 flaring event. We used the code described in \citet{Cerruti15} that computes the emission at equilibrium by a spherical plasmoid in the jet, parametrized by its radius $R'$, and Doppler factor $D$, filled with a homogeneous magnetic field $B'$. The code assumes that primary populations of electrons and protons are present in the emitting region. Hadronic interactions \revC{spawn} secondary particles (photons from $\pi^0$ decay and electrons/positrons from $\pi^\pm$ decay and Bethe-Heitler pair production) in the emitting region: these trigger electron-positron pair cascades that radiate via both synchrotron and inverse Compton scattering. Neutrinos instead leave the emitting region with \revC{a negligible probability of} interactions. Protons interact with both the internal photon fields (synchrotron and inverse Compton emission by primary electrons) and with external ones that are parameterized by a gray-body spectrum with temperature $T'=2\times 10^6\ $K, commensurate with the mean photon energy quoted in Eq.~\ref{eq:16}. 

Primary protons are \revC{assumed to be described} by a simple power-law energy distribution with an exponential cut-off, while primary electrons are described by a broken power law with an exponential cut-off. \revHESSc{This choice is motivated by the fact that primary electrons are certainly cooled due to synchrotron radiation, while primary protons are supposed here to be not cooled: this hypothesis is confirmed a posteriori by comparing all relevant timescales.} The primary electron distribution is thus defined by six free parameters (the two \revHESS{power-law indices} \qf{$\alpha_{e,1}$ and $\alpha_{e,2}$}, the minimum/break/maximum Lorentz factors \qf{$\gamma_{\text{min,e}}', \gamma_{\text{br,e}}'$, and $\gamma_{\text{max,e}}'$}, and the normalization \qf{number density $n_\text{e}'$}), and the primary proton distribution by an additional four (the index \qf{$\alpha_\text{p}$}, minimum/maximum Lorentz factors \qf{$\gamma_{\text{min, p}}'$ and $\gamma_{\text{max, p}}'$}, and the normalization \qf{number density $n_\text{p}'$}). 

\begin{figure*}[hb]
\plotone{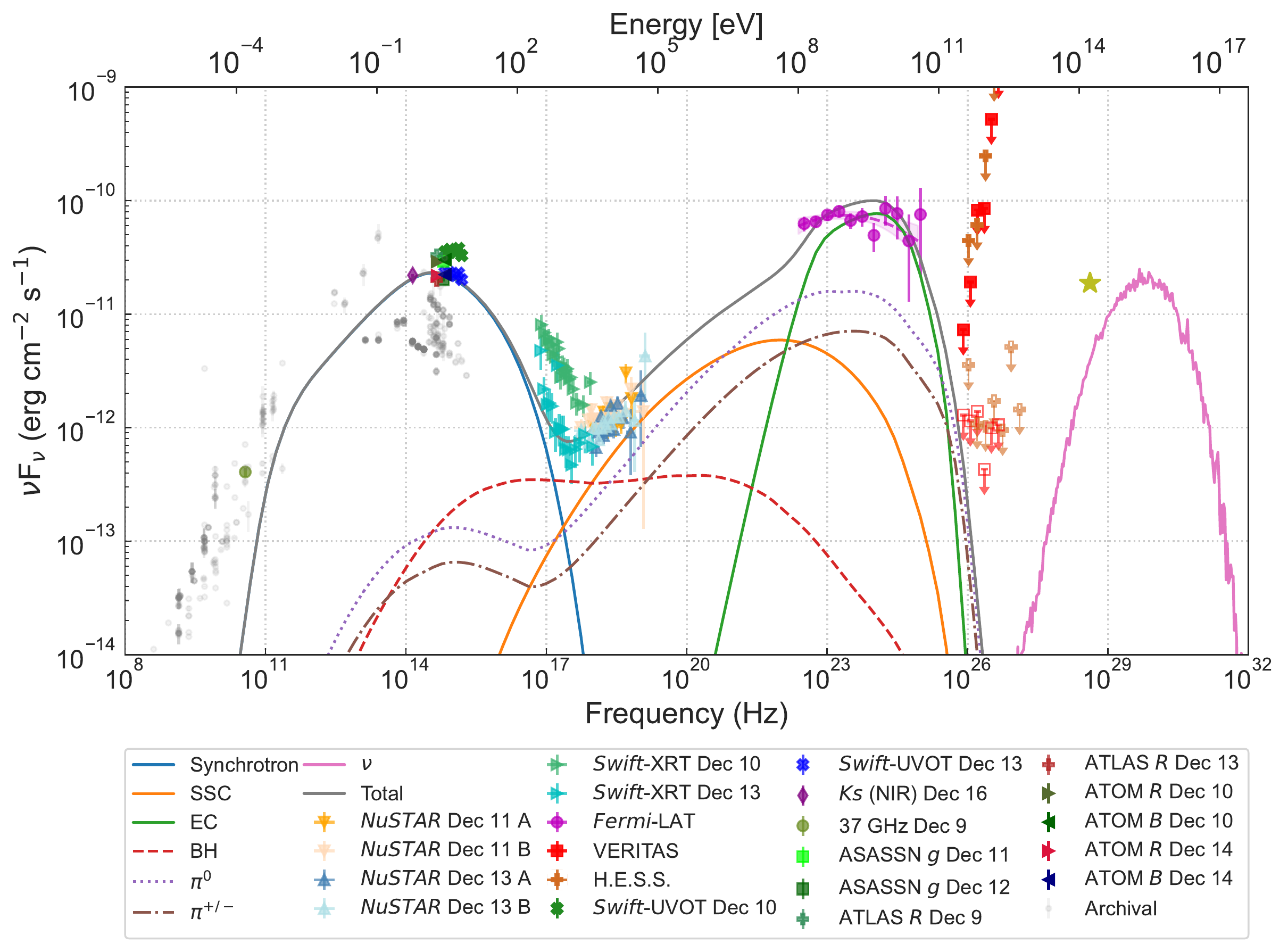}
\caption{The lepto-hadronic model with an external photon field, as described in Section~\ref{subsec:numericalSED}. The solid blue, orange, and green lines are the synchrotron, SSC, and external Compton components of the leptonic emission from the primary electrons, respectively; the dashed red line is the combined synchrotron and inverse-Compton emission from Bethe-Heitler pairs; the dotted and dash-dotted lines are the neutral and charged pion cascade components, respectively; the solid gray line is the sum of all electromagnetic components; and the solid pink line above a few TeV is the neutrino emission. The yellow star marks the nominal flux of 1.5 170-TeV neutrinos per year using an effective area of $30\ \mathrm{m}^2$ (see Subsections~\ref{subsubsec:neu1} and \ref{subsec:numericalSED}) to guide the eye. 
\label{fig:sed_hadro}}
\end{figure*}

Together with the three parameters \qf{($B'$, $D$, and $R'$)} describing the emitting region, and the normalization of the external photon density \qf{$U_{bbody}'$}, the total amount of free parameters of the model is 14. 
\revB{Instead of fitting a model with many free parameters, we aimed at} a solution that describes the photon SED and maximizes the neutrino output. \revB{The indices of the injected primary particles are fixed} to $\alpha_{e,1} = \alpha_p = 2.0$, while that of the cooled electrons is fixed to $3.0$ as expected from simple synchrotron cooling. The Doppler factor of the emitting region is fixed at $30$, a typical value for blazars. The other free parameters are manually optimized to reproduce the SED via a dominant SSC and EIC component in the \g{} band, and a sub-dominant hadronic component. The latter cannot be arbitrarily luminous; it is mainly constrained by the Bethe-Heitler component that emerges in the X-ray band. Neutrino rates are computed using the effective area \qf{$A_\mathrm{eff}\approx 30\ \mathrm{m}^2$}. In Fig. \ref{fig:sed_hadro} we show the \mpon{SED resulting from the} lepto-hadronic modeling, with both the photon and neutrino components. 

A detailed list of model parameters is provided in Table \ref{tab:hadronic_model}. The expected neutrino rate is \qfn{$\sim1.5$} \revHESSc{events} per year, or \qfn{$0.125$} events in a 30-day activity period, \qf{marginally} consistent with the detection of a single event by IceCube. The power of the jet needed to produce the amount of photons and neutrinos is $2 \times 10^{48}$ erg s$^{-1}$ -- roughly 
\qfn{20} times higher than the Eddington luminosity, \mc{of a supermassive black hole with a mass of $10^9\ M_\odot$}, 
\mpon{but it is} 
very dependent on the choice of $\alpha_\text{p}$ and $\gamma_\text{p, min}$. The values assumed here, $2.0$ and $1$, respectively, are conservative with respect to the total jet power. A harder proton injection or a higher minimum Lorentz factor of protons significantly reduces the energetic requirement \revC{and would bring the particle energy density closer to that of the magnetic field}. As an example, an equivalent SED modeling, with similar neutrino rates, is possible assuming $\alpha_p = 1.9$ and  $L_{jet} = 8.5\times10^{47}$~erg~s$^{-1}$.


\begin{table}
    \centering
    {\caption{\revHESS{Parameters for the lepto-hadronic modeling of the SED.}}
    \label{tab:hadronic_model}
        \begin{tabular}{ c | c  }     
        \toprule
            \revHESSc{\sc{Parameter\footnote{The quantities flagged with an asterisk are derived quantities and not model parameters. The luminosity of the emitting region has been calculated as $L=2\pi R^\prime c (D/2)^2(U^\prime_\text{B}+U^\prime_\text{e}+U^\prime_\text{p})$,where $U^\prime_\text{B}$, $U^\prime_\text{e}$, and $U^\prime_\text{p}$, are the energy densities of the magnetic field, the electrons, and the protons, respectively.}}}& \revHESSc{\sc{Value}}\\
            \midrule
            z & 0.45\\
            $D$ & 30\\
            $B^\prime$ [G] & 0.3 \\
             $R^\prime$ [cm] & $2.1 \times 10^{16}$ \\
             $n_\text{e}^\prime$ [1/cm$^{3}$] & $4\times10^3$ \\
             $n_\text{e}^\prime$/$n_\text{p}'$ & 1.0 \\
             $\alpha_\text{e,1}$ & 2 \\
             $\alpha_\text{e,2}$ & 3 \\

             $\gamma_{\text{min,e}}^\prime$ & 100  \\
             $\gamma_{\text{br,e}}^\prime$ & $3.5\times10^3$ \\
             $\gamma_{\text{max,e}}^\prime$  & $1.3\times 10^4$ \\
             $ \alpha_\text{p}$ & 2.0 \\
             $\gamma_{\text{min, p}}^\prime$ & 1\\
             $\gamma_{\text{max, p}}^\prime$ & $1 \times 10^{7}$ \\
             $T_{bbody}^\prime$ [K] & $2\times 10^6$\\
             $U_{bbody}^\prime$ [erg/cm$^{3}$] & $0.16$\\
            &\\
            $^\star U^\prime_\text{e}$ [erg/cm$^{3}$]  & $1.2 \times 10^{-2}$ \\
            $^\star U^\prime_\text{B}$ [erg/cm$^{3}$] & $3.6 \times 10^{-3}$\\
            $^\star U^\prime_\text{p}$ [erg/cm$^{3}$] & $93$ \\
            $^\star U^\prime_\text{e}$ / $U^\prime_\text{B}$ & $3.5$ \\
            $^\star U^\prime_\text{p}$ / $U^\prime_\text{B}$ & $2.6 \times 10^{4}$ \\
            $^\star L_{\text{jet}}$ [erg/s] & $2.0 \times 10^{48}$ \\
            $^\star \nu_\mathrm{BRONZE}$~[year$^{-1}$] & $1.49$ \\ 
            \bottomrule   
        \end{tabular}
        }
\end{table}

\section{Summary}

A few days after the neutrino event \revHESSc{\neu{}}, detected by IceCube with an energy $E_\nu\approx 171$ TeV on December 8, 2021, the blazar PKS~0735+178 was observed at an elevated flux state from optical to \g{} bands, reaching the highest soft X-ray and GeV \g{} fluxes among all measured values since 2007. \revHESSc{PKS~0735+178} is located 2.2$^\circ$ away from the best-fit position of the IceCube event, immediately outside \revHESS{the 90\%} error region (2.13$^\circ$). Its soft X-ray flux exhibited fast variability on a one-day timescale. The active flux state of the source lasted \revHESSc{from} roughly two weeks to a month. 

The broadband SED of PKS~0735+178 near the time of the IceCube event was well measured by extensive follow-up observations across the electromagnetic spectrum. \revHESS{In particular, the X-ray data from \swift{}-XRT and \nustar{} characterize the tail of the synchrotron SED peak and the onset of the high-energy SED peak.  The \g{} data from \fermi{}-LAT, VERITAS and \hess{} require a spectral cut-off \mpon{near} 100 GeV, after taking into account the EBL absorption.}

We analytically demonstrated that the observed SED of the blazar, \revD{especially the \g{} cutoff in the TeV data,} constitutes a challenge to a simple one-zone SSC model, unless, \revB{in the unlikely scenario, the synchrotron peak extends into the far-UV band \revC{that the observations do not cover}}. 
\revHESSc{Alternatively, it can be explained by an SSC/EC scenario, which naturally provides the observed 100 GeV cut-off through the Klein-Nishina effect and $\gamma$-$\gamma$ pair absorption. It could also be explained by a lepto-hadronic mechanism with an external photon field, which marginally agrees with the Eddington limit on the \revC{jet} luminosity and the observed IceCube neutrino rate.}

We presented a numerical lepto-hadronic model with external target photons that is consistent with the observed SED and marginally consistent with the neutrino event. In this model, the electromagnetic emission is dominated by the leptonic components, while the sub-dominant hadronic components are constrained by the observed X-ray spectrum. The jet power is \mpon{significantly} higher than the Eddington luminosity, unless the spectrum of the parent protons is \mc{harder than 2.0}. The expected total neutrino rate is 1.5 per year, or 0.125 per month. 

Some challenges in searching for neutrino-emitting blazars are common in the cases of PKS~0735+178 and TXS~0506+056: the limited localization precision of the IceCube Observatory and the large number of \g{} blazars as potential counterparts make the association between neutrino events and \g{} blazars difficult; the electromagnetic emission can often be explained by leptonic models alone without the need for \revHESSc{a hadronic component}; the jet power and the proton luminosity required in lepto-hadronic models are often too high, exceeding the Eddington limit, for a short period of activity. \revB{These challenges can only be addressed with additional data and continued searches for astrophysical neutrino emitters}, including follow-up observations of flaring blazars in temporal and spatial coincidence with astrophysical neutrino events. Such follow-up efforts remain a focus in the multi-messenger community \citep[see e.g., ][]{XRT_IC220303A, LAT_IC220822A, 2022icrc.confE.960T}. 



\begin{acknowledgments}
This research is supported by grants from the U.S. Department of Energy Office of Science, the U.S. National Science Foundation and the Smithsonian Institution, by NSERC in Canada, and by the Helmholtz Association in Germany. This research used resources provided by the Open Science Grid, which is supported by the National Science Foundation and the U.S. Department of Energy's Office of Science, and resources of the National Energy Research Scientific Computing Center (NERSC), a U.S. Department of Energy Office of Science User Facility operated under Contract No. DE-AC02-05CH11231. We acknowledge the excellent work of the technical support staff at the Fred Lawrence Whipple Observatory and at the collaborating institutions in the construction and operation of the instrument. 
\end{acknowledgments}
\begin{acknowledgments}
This work was supported by NASA grants 80NSSC22K0573, \revB{80NSSC22K1515, 80NSSC22K0950, 80NSSC20K1587, 80NSSC20K1494,} and NSF grant PHY-1806554. 
\end{acknowledgments}
\begin{acknowledgments}
The support of the Namibian authorities and of the University of
Namibia in facilitating the construction and operation of \hess{}
is gratefully acknowledged, as is the support by the German
Ministry for Education and Research (BMBF), the Max Planck Society,
the German Research Foundation (DFG), the Helmholtz Association,
the Alexander von Humboldt Foundation, the French Ministry of
Higher Education, Research and Innovation, the Centre National de
la Recherche Scientifique (CNRS/IN2P3 and CNRS/INSU), the
Commissariat à l’énergie atomique et aux énergies alternatives
(CEA), the U.K. Science and Technology Facilities Council (STFC),
the Irish Research Council (IRC) and the Science Foundation Ireland
(SFI), the Knut and Alice Wallenberg Foundation, the Polish
Ministry of Education and Science, agreement no. 2021/WK/06, the
South African Department of Science and Technology and National
Research Foundation, the University of Namibia, the National
Commission on Research, Science \& Technology of Namibia (NCRST),
the Austrian Federal Ministry of Education, Science and Research
and the Austrian Science Fund (FWF), the Australian Research
Council (ARC), the Japan Society for the Promotion of Science, the
University of Amsterdam and the Science Committee of Armenia grant
21AG-1C085. We appreciate the excellent work of the technical
support staff in Berlin, Zeuthen, Heidelberg, Palaiseau, Paris,
Saclay, Tübingen and in Namibia in the construction and operation
of the equipment. This work benefited from services provided by the
\hess{} Virtual Organisation, supported by the national resource
providers of the EGI Federation.
\end{acknowledgments}
\begin{acknowledgments}
This work made use of data supplied by the UK \swift{} Science Data Centre at the University of Leicester.
\end{acknowledgments}
\begin{acknowledgments}
This work has made use of data from the Asteroid Terrestrial-impact Last Alert
System (ATLAS) project. The Asteroid Terrestrial-impact Last Alert System
(ATLAS) project is primarily funded to search for near earth asteroids through
NASA grants NN12AR55G, 80NSSC18K0284, and 80NSSC18K1575; byproducts of
the NEO search include images and catalogs from the survey area. This work was
partially funded by Kepler/K2 grant J1944/80NSSC19K0112 and HST GO-15889,
and STFC grants ST/T000198/1 and ST/S006109/1. The ATLAS science products
have been made possible through the contributions of the University of Hawaii
Institute for Astronomy, the Queen’s University Belfast, the Space Telescope
Science Institute, the South African Astronomical Observatory, and The
Millennium Institute of Astrophysics (MAS), Chile.
\end{acknowledgments}

%

\facilities{VERITAS, H.E.S.S., \revHESSc{\fermi{}-LAT, \nustar{}, \swift{}-XRT, \swift{}-UVOT, ASAS-SN, ATLAS, NOT, OVRO}} 


\software{Astropy \citep{Astropy13,Astropy2018},  
          NumPy \citep{numpy11},
          Matplotlib \citep{Hunter07},
          SciPy \citep{SciPy},
          VEGAS \citep{Cogan08}, Eventdisplay \citep{Maier17}, Model Analysis \citep[][]{hess_denaurois_2009},
          Fermitools \citep{Fermitools2019}, HEAsoft \citep{HEAsoft2014}
          }


\bibliography{qi_bib_all}{}
\bibliographystyle{aasjournal}

\allauthors



\end{document}